\documentclass[twocolumn,twocolappendix]{aastex63}
\usepackage{amsmath}

\newcommand\teff{$T_{\rm eff}$}
\newcommand\prot{$P_{\rm rot}$}

\defcitealias{kounkel2018a}{Paper I}
\usepackage{listings}

\begin{document}
\title{Using rapid rotators as tracers of multiplicity statistics as a function of stellar density}

\author{Priyanka Cingirikonda}
\affil{Creekside High School, 100 Knights Ln, St Johns, FL 32259}
\affil{Department of Physics and Astronomy, University of North Florida, 1 UNF Dr, Jacksonville, FL, 32224, USA}
\author[0000-0002-5365-1267]{Marina Kounkel}
\affil{Department of Physics and Astronomy, University of North Florida, 1 UNF Dr, Jacksonville, FL, 32224, USA}
\email{marina.kounkel@unf.edu}
\author[0000-0002-1650-2764]{Joseph Mullen}
\affil{Department of Physics and Astronomy, University of North Florida, 1 UNF Dr, Jacksonville, FL, 32224, USA}
\affil{Department of Physics and Astronomy, Vanderbilt University, VU Station 1807, Nashville, TN 37235, USA}

\begin{abstract}

Recent works have identified that rapidly rotating stars are predominantly binaries with separations of a few to a few tenths of au. This is a crucial range of separation that is often inaccessible to searches of binary stars, providing a unique opportunity to examine their statistical properties. In particular, we have performed an analysis of rapid rotators in young moving groups. We examined their fraction as a function of the stellar density of the population in which they are found. We find that there is a deficit of rapid rotators in dense clusters such as the Orion Nebula in comparison to the more diffuse parts of the Orion Complex, as intracluster interactions with neghboring stars likely dissolve binaries with intermediate separations before they had a chance to fully form. In contrast, in older populations with an age of $\sim100$ Myr, mass segregation redistributes binaries relative to single stars, thus in such older regions, rapid rotators are predominantly found in the regions of higher stellar density. This work sheds light on both the conditions that lead to the formation of binary stars and their dynamical evolution. 
\end{abstract}


\section{Introduction}

Approximately half of solar-type stars are found in binary systems, and they can be found at a wide range of separations from $<$0.1 au, to $>$100,000 au, with most common separations of 30--40 au \citep{raghavan2010}. Depending on their separation, different methods are needed to search for them - systems with short separations can be found using kinematics, e.g., searching for radial velocity variability, searching for acceleration in proper motions, etc. However, this technique becomes inefficient at the orbital periods longer than several years due to increasingly subtler change in the kinematics that takes longer to manifest \citep[e.g.,][]{price-whelan2020}. Wide binaries, on the other hand, can be found through high resolution imaging, where both sources can be resolved from one another, as long as it is possible to confirm that they are not chance alignments of unrelated stars \citep[e.g.,][]{el-badry2021}. 

For nearby stars, where small angular separations between resolved systems translate to spatial separations on the order of a few au, both methods can be seamlessly stitched together to provide a mostly complete census of binaries across a full distribution of orbital periods. On the other hand, even at distances to stars of just a few hundred pc, it becomes increasingly harder to close the gap between these two approaches, i.e., binaries can be characterized at separations of less than a few au, or greater than several 100s of au, but not at the separations where binaries are most common.

This becomes particularly problematic when it comes to understanding multiplicity in nearby star forming regions. It is believed that the bulk of properties of stars pertaining to multiplicity, including the overall frequency of binaries and the distribution of separations \citep[e.g.][]{offner2023}, should be set very early on, with most processes governing it finishing in $\lesssim$1 Myr -- the timescale comparable to the star assembling its final mass, and the companion migrating from its initial radius at formation to settle at its stable orbit. And, it is believed that upon dissolving all of the nearby star forming regions, the overall distribution of multiples produced by them should mirror the field \citep{kroupa2001,marks2012}.

However, the closest major star forming region to us is located $>100$ pc away (e.g., Taurus, Sco Cen), and the bulk of particularly massive nearby populations (e.g., Orion, Vela, Perseus) are found at distances of 300--400 pc. This makes finding a complete census of multiples a challenge.

It has long been observed that wide binaries exhibit different properties depending on the star forming region in question: e.g., in diffuse populations such as Taurus, the fraction of multiples at separations $>$100 au is well in excess of what is observed in the field \citep[and references therein]{kraus2011}. On the other hand, dense clusters, such as the Orion Nebula, were deficient in very wide binaries \citep{scally1999,reipurth2007,jerabkova2019}. These differences were typically attributed to the difference in stellar density. Through dynamical interactions with the neighboring stars, wide systems get disrupted and dissolve in denser star forming environments, but they can survive in more diffuse regions \cite{marks2012}. Upon being released into the field, the mixture of stars formed in different environments would reconstruct the field distribution. This was supported by studies of spectroscopic binaries in these regions showing no obvious differences between different regions \cite{kounkel2019}, as at these close separations, the effect from neighboring stars is expected to be mostly negligible.

However, the behavior of binaries at intermediate separations remained a puzzle. Although some high resolution studies were able to push the limit of resolved binaries down to $\sim10$ au \citep{kraus2011,duchene2018,de-furio2019}, the coverage remains sparse, and the overall selection biases that are intrinsic to each survey make a direct comparison between these regions difficult, and any initially apparent differences between them remain inconclusive \citep{eastlund2025}.

Recently, though, a new method was identified that may allow to examine statistics of binaries at these intermediate separations, albeit indirectly, through observing stellar rotation period. Over the course of their lives, low mass stars with convective envelopes exhibit strong magnetic fields that exert torque onto a star, incrementally slowing down its rotation over time. At a given age, a star of a given mass is likely to have a relatively narrow distribution of rotational periods \citep[e.g.,][]{curtis2020}. However, there exists a population of rapid rotators -- stars that have rotational periods significantly faster than what is expected for their age. It has been demonstrated that these rapid rotators appear to be predominantly binaries, that multiplicity can cause this rapid rotation in stars that are younger than 1 Myr, and that it is primarily driven by unresolved but not spectroscopic binaries, i.e., systems with separation of a few to a few tens of au \citep{kounkel2022a,kounkel2023a}. Although the exact mechanisms behind it are still not fully known, it has been suggested that a combination of outflows being inhibited by a companion at intermediate range of separations \citep[leading to their angular momentum loss being less efficient than for single stars,][]{kuruwita2017} and mutual torques that both stars would experience \citep{kuruwita2024} could lead to rapid rotation for these binaries.

A signature of rapid rotation is less definitive than observation of resolved systems, or observations of kinematics -- e.g., it sheds no light on the mass ratio, on the exact separation between the two stars, and even the identified sources can only be considered as candidate multiples, requiring external confirmation of their nature. Nonetheless, through examining the population of rapid rotators at different ages and in different environments, it may be possible to shed light at the properties of binaries with intermediate separations that are difficult to access through other means. 

In this paper, we present a study of young rapid rotators in 6 different nearby populations with ages between $<$5 Myr and 150 Myr that span a wide range of densities within each region. In Section \ref{sec:data} we present the dataset used in this analysis. In Section \ref{sec:results} we perform the analysis. In Section \ref{sec:discussion} we discuss the implications of our observations, and in Section \ref{sec:conclusions} we conclude our results.

\section{Data}\label{sec:data}

\subsection{Initial selection}

\citet{kounkel2022a} have obtained rotational periods (\prot) of $\sim$100,000 stars through analyzing light curves from Transiting Exoplanet Survey Satellite (TESS). The analyzed sources have been identified as members of nearby ($\lesssim$ 500 pc) young moving groups with ages of $\lesssim$1 Gyr. These moving groups were identified through performing hierarchical clustering \citep{kounkel2019a,kounkel2020} based on astrometry provided by Gaia DR2 \citep{gaia-collaboration2018}.

With this, the total census of sources we included in our sample consisted of the following criteria

\begin{itemize}
    \item Measurements of \prot\ from \citet{kounkel2022a}; measurements of stellar mass and radius from TESS Input Catalog \citep{stassun2019} to provide a defined measurement of angular momentum ($L \equiv \frac{4\pi MR^2}{5P_{rot}}$ assuming solid body rotation of a sphere)
    \item Extinction corrected color [$G_{\rm BP}-G_{\rm BP}$]$>$ 0.7 mag, to ensure the selected sources are below the Kraft break, i.e., they have convective envelopes, to minimize contamination from variables that are not rotating
\end{itemize}

In conjunction with measuring rotational periods, \citet{kounkel2022a} have performed a calibration for gyrochronology, obtaining a transformation between angular momentum $L$ and age as a function of \teff. This transformation is valid for slow rotators, i.e., predominantly single stars. 

\begin{figure}
\epsscale{1.2}
\plotone{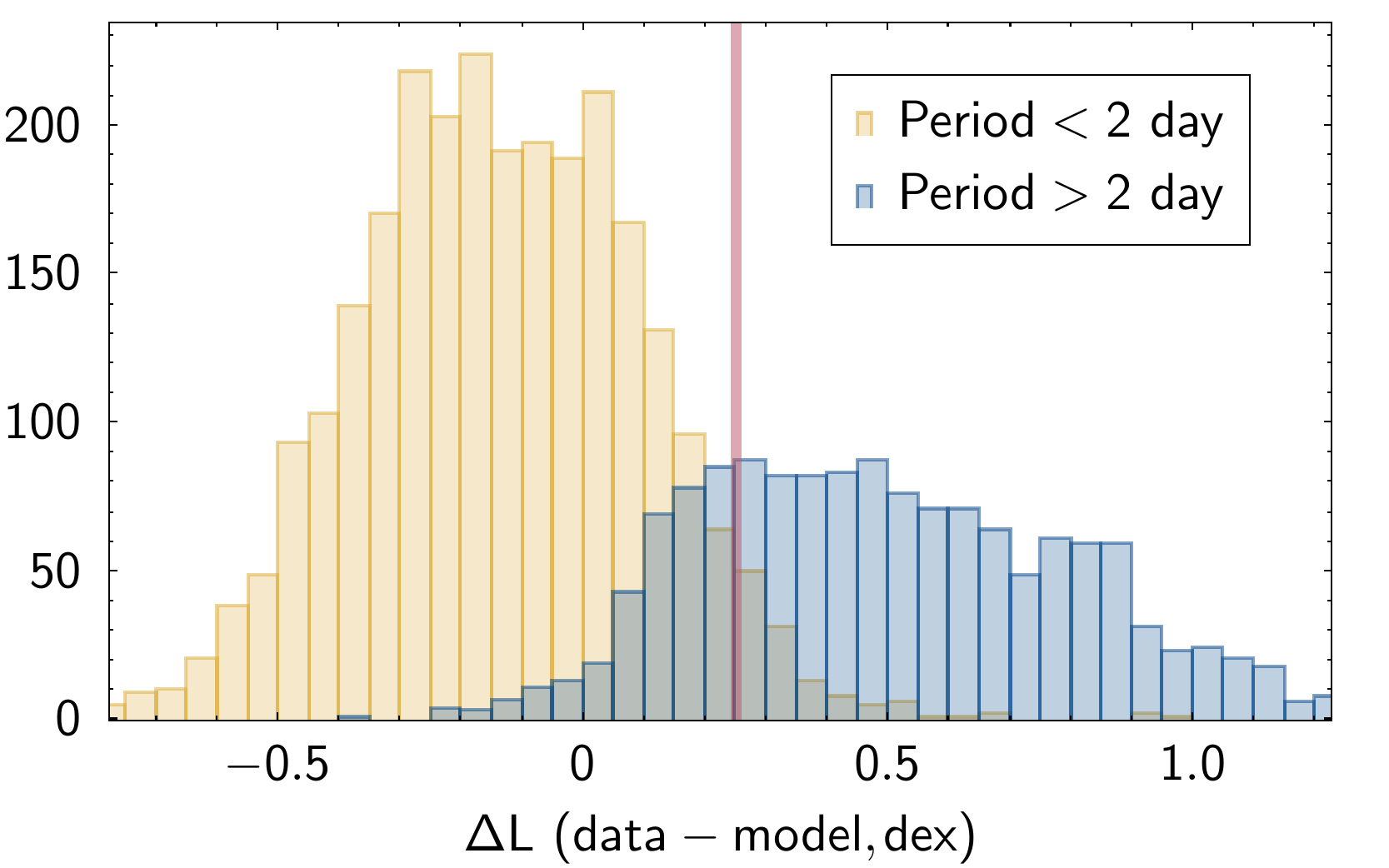}
\caption{Comparison of angular momentum difference of stars in Orion between the data and the model, highlighting the sources that have been previously identified as rapid rotators in this region.}
\label{fig:ori}
\end{figure}

We use this relation to identify rapid rotators, which should have significantly higher $L$ than the model. \citet{kounkel2023a} have found that in the Orion Complex, rapid rotators are the sources with \prot$<$2 days. Converting rotational periods to angular momentum, this selection translates to an offset in $\log L$ of 0.25 dex relative to the model (Figure \ref{fig:ori}). Thus, as a subset of the above sample, rapid rotators were selected via
\begin{itemize}
    \item $\log L$-$\log L_{\rm model}>0.25$ dex
\end{itemize}

\subsection{Contamination}

TESS has a very large pixel scale, $\sim$21''. As a result, in dense clusters, periodicity in the light curves of individual stars can actually be contamination from the nearby sources. I.e., if a particular source has a particular \prot, the same \prot\ can manifest in the nearby sources that surround it.

On average, as has been demonstrated by \citet{kounkel2022a}, with sufficiently large sample size, such contamination can be ignored. However, the study presented in this paper seeks to examine differences between dense and diffuse regions, and contamination is expected to be significantly higher in denser regions.

To minimize the effects of contamination, we apply \texttt{TESS\_localize} \citep{higgins2023}, which enables de-blending flux from multiple sources and locating the source of periodicity to better than 1/5 of a pixel and identify a source in Gaia DR3 catalog that has the highest likelihood of being the source of variability with a specified period, although it may also struggle to perform in very dense regions (such as e.g., Orion Nebula).

Over the course of the analysis presented in the paper we examine the full sample, as the greater quantity of sources enables larger signal-to-noise. However, to verify our findings, we also perform an analysis of the subset of sources where their \prot\ has been explicitly verified by \texttt{TESS\_localize} to have the highest probability of originating from the source in question.

\subsection{Stellar density}

In addition to identifying a subset of rapid rotators, we estimate stellar density within each population through measuring distance to the 4th nearest neighbor, similarly to \citet{kounkel2016a}. The search for the 4th nearest neighbor was conducted in 3d space, obtaining X, Y, \& Z coordinates for all the stars from their position on the sky and parallax (which was available for all of the stars in the sample, and all of the regions are sufficiently nearby to provide good precision in distance, typically on the order of a few pc). A 3-dimensional approach was needed due to the fact that many extended populations span on the order of 100 pc in width and depth, significantly larger than the typical uncertainties in the parallax, thus calculating 4th nearest neighbor distance in projection would have introduced significant distortions.

In examining the frequency of appearance of rapid rotators, comparing distinct populations relative to one another may introduce significant biases. For example, the maximum \prot\ that can be reliably measured by TESS is $\sim$12 days, i.e., half of a TESS's sector, as multiple sectors are difficult to stitch together. While such a limit is sufficient to identify all rotators in the youngest populations with the age of a few Myr, and it should be sufficient to find all of the rapid rotators, at some point, slowly rotating K-dwarfs would be missing from the sample.

This would potentially introduce age dependent biases to the underlying multiplicity fraction that may be inferred from a given population. Thus, we do not attempt to compare different groups relative to one another directly. Instead, we analyze distribution of rapid rotators within each single moving group individually, as both dense and diffuse parts of a single populations would have comparable distribution of ages to within a few Myr, and thus the aforementioned biases would affect different parts of the same population similarly.

Not all moving groups have a sufficiently large enough number of sources to provide reliable statistics, particularly as the overall census of sources in this group would also need to be split into different bins of 4th nearest neighbor distance to examine variations attributable to the stellar density.

In total, we have identified 6 moving groups cataloged in \citet{kounkel2020} that had at least 20 identified rapid rotators in several discrete density bins. These groups include

\begin{itemize}
    \item Theia 13 (Orion Complex), age $<$8 Myr (1199 rapid rotators of 4066 stars)
    \item Theia 22 (Vela OB2), age $\sim$10 Myr (385 rapid rotators of 917 stars)
    \item Theia 43 (A part of Sco Cen OB association extended from Corona Australis to Lower Centaurus Crux. Does not include Upper Sco and Upper Centaurus Lupus - in part due to the former being in the gap between TESS sectors and not observed in the first three years of observations), age $\sim$15 Myr (343 rapid rotators of 919 stars)
    \item Theia 74 (includes NGC 2547), age $\sim$30 Myr (216 rapid rotators of 595 stars)
    \item Theia 143 (includes Trumpler 10), age $\sim$60 Myr (142 rapid rotators of 453 stars)
    \item Theia 613 (includes NGC 2516), age $\sim$150 Myr (405 rapid rotators of 1011 stars)
\end{itemize}

\section{Results}\label{sec:results}

\begin{figure*}
\epsscale{1.2}
\plotone{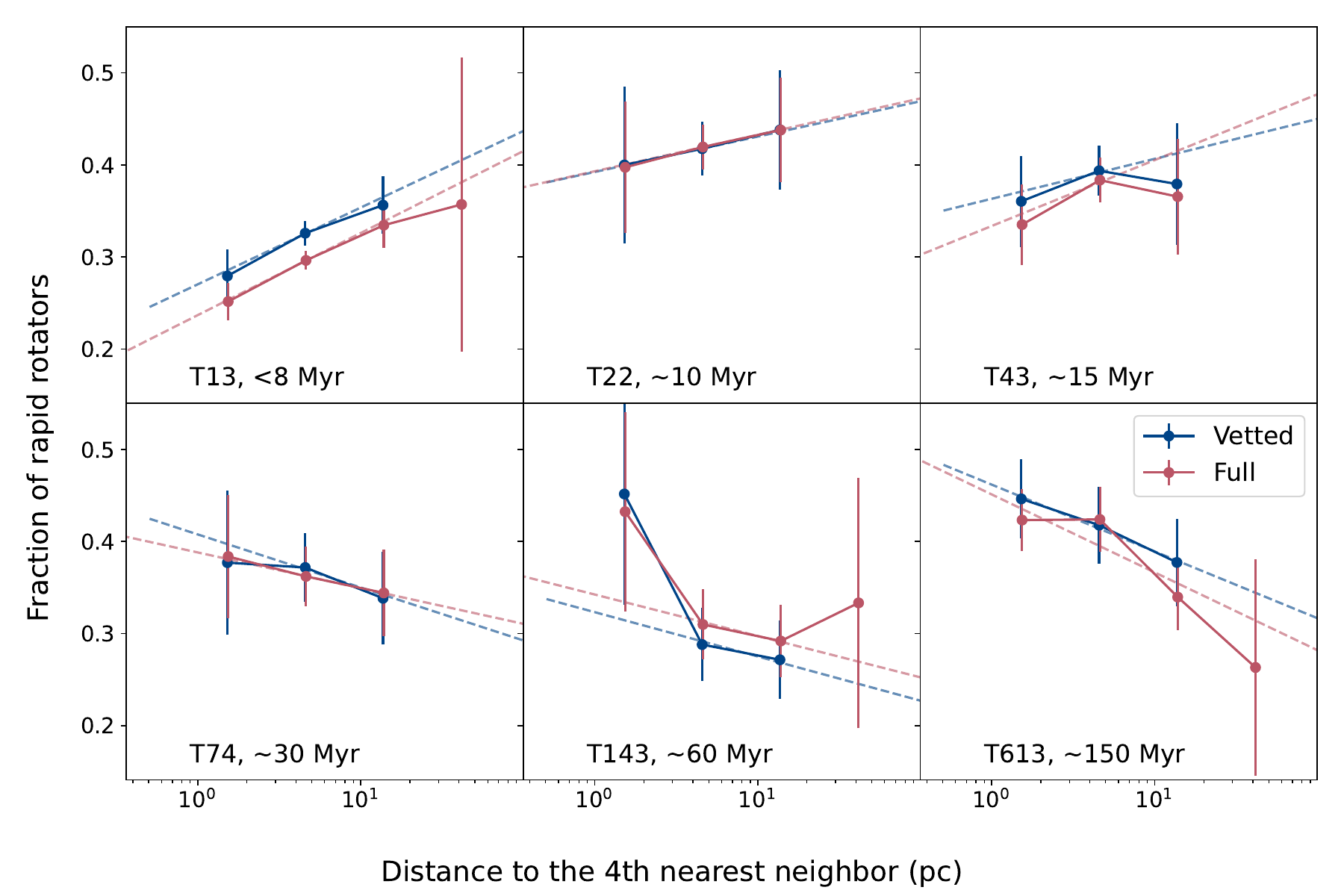}
\caption{Fraction of rapid rotators as function of fourth nearest neighbor distance in each of the moving groups used in the analysis. Smaller distance to the fourth nearest neighbor translate to higher density. The best fitted slope is shown with the dashed line. Only bins with at least 5 rapid rotators are shown. Red line shows the full sample, blue line shows only the sample where periods have been confirmed with \texttt{TESS\_localize}. The plot shows younger populations having a more positive slope, i.e., more rapid rotators are found in the more diffuse regions. On the other hand, older populations have a more negative slope, i.e., more rapid rotators are found in denser parts of the population.
\label{fig:fractions}}
\end{figure*}

We have separated the members with convective envelopes for which rotational periods are available for all 6 identified populations into multiple logarithmically spaced bins of varying density that has been inferred by the 3d distance 4th nearest neighbor. In each of these bins we have calculated the fraction of rapid rotators to the total number of sources. Uncertainties in the fractions have been calculated using Gaussian $\sqrt{N}$ approximation. We have performed this analysis separately for the full sample, and the sample that has been vetted by \texttt{TESS\_localize}. This fraction of rapid rotators is shown in Figure \ref{fig:fractions}.

The fraction of rapid rotators appears to typically vary between 0.2 to 0.4, but as has been previously mentioned, due to different biases involved in the selection of members of different populations, it is not trivial to compare the overall fractions between different populations directly.

Of greater interest is to examine trends in any particular population. We find that in the younger moving groups (Theia 13, 22, 43), the fraction of rapid rotators is inversely proportional with density - rapid rotators are less common in regions with small distance to the fourth nearest neighbor, and rapid rotators become more common in more diffuse regions.

On the other hand, in older populations (Theia 74, 143, 613), the trend appears to reverse: more rapid rotators are found in denser regions, and fewer in more diffuse regions. Across all of the populations, there is no substantial difference between the full sample and the sample vetted by \texttt{TESS\_localize} - any contamination of periodicity from the nearby sources appears to not affect the overall statistics in a significant manner.

The overall trend of the fraction of rapid rotators as a function of 4th nearest neighbor distance appears to follow a log-linear trend. We have performed least squares fitting to find the slope of these trends. They are visualized as a function of age in Figure \ref{fig:slope}.




\section{Discussion}\label{sec:discussion}

\subsection{Implication on the initial likelihood of forming binaries with intermediate separations in different environments}

Youngest moving groups appear to have significantly fewer rapid rotators in denser regions than in the more diffuse regions. Given the age of these populations of just a few Myr, it is likely that this distribution is mostly primordial, and has experienced only minimal evolution.

If we assume that these rapid rotators well trace the underlying distribution of binaries with separations of a few to several tens of au, this appears to suggest that these intermediate separations binaries are less common in denser environments, consistent with the assumption of dynamical evolution \citep{kroupa2001,marks2012}. The difference between the dense and diffuse regions appears to be particularly stark in the youngest region, the Orion Complex (T13), where the stellar density varies from $>10^4$ stars pc$^{-3}$ \citep{hillenbrand1998} at the center of the most massive nearby cluster, the Orion Nebula -- to the density approaching the field near the outskirts.

\citep{duchene2018} have previously argued that although dynamical processing of wide binaries in the Orion Nebula is to be expected, systems tighter than 60 au should be ``pristine'' even in a cluster with such a high stellar density, unaffected by the random encounters to unbind them. Since we do not know what are the separations or mass ratios of the candidate multiples that are identified on the basis of rotation, we are unable to test this directly with this sample.

It is commonly accepted that systems with the separations of few tens of au should be ``pristine'', i.e., not affected by dynamical encounters with other stars, even in dense clusters thus the multiplicity fraction at these separations should be independent of stellar density \citep[e.g.,][]{duchene2018}. However, it is also believed that early on, protostellar and prestellar multiples, particularly those that formed from core fragmentation, are likely to form widely separated. Over the course of $\ll$1 Myr they will either harden to closer separations or dissolve as they are affected by the torques from both the protoplanetary disk and from the other neighbors that are found in the same cluster \citep{goodwin2007,connelley2008,ramirez-tannus2021,kuruwita2022}. Thus many of the systems that will end up with separations of $<60$ au even at a young age of 1 Myr could have began their lives at separations of $>1000$ au. This means that although present day intermediate separation binaries that are found in young clusters are no longer likely to be affected by the dynamical encounters, there can nonetheless be a deficit of such multiples, because many of them could have dissolved before they got a chance to fully harden.

\subsection{Evolution of the distribution of binaries with intermediate separations with age}

\begin{figure}
\epsscale{1.2}
\plotone{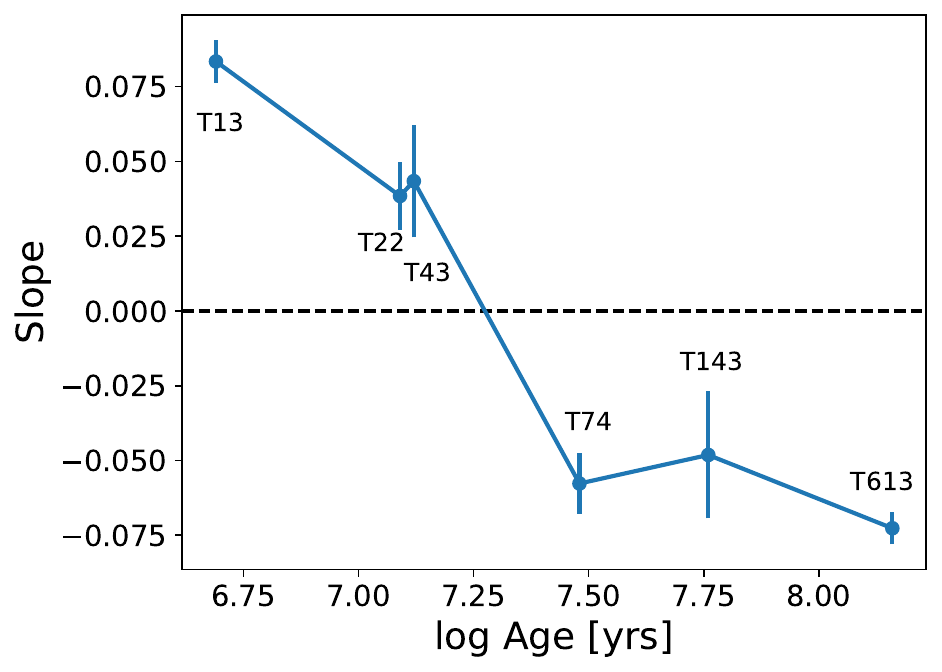}
\caption{Evolution of slopes of the fraction of rapid rotators as a function of density shown in Figure \ref{fig:fractions} relative to the age of each population.
\label{fig:slope}}
\end{figure}

In contrast to the younger populations, older moving groups show an opposite trend: binary candidates that are traced by rapid rotators appear to be more heavily concentrated in the regions with higher stellar density. Moreover, there does appear to be a steady progression of the fraction of rapid rotators (and thus the inferred multiplicity fraction of binaries with intermediate separations) with age (Figure \ref{fig:slope}). In the youngest regions, rapid rotators are heavily concentrated in the more diffuse outskirts, but even after 10 Myr, the slope of the distribution of rapid rotators with respect to stellar density becomes shallower. At 20 Myr, there appears to be a turnover where they would be equally commonly located in both dense and diffuse parts of the moving group, and at the ages beyond that, they seem to be more common in the centers of the denser clusters compared to the more diffuse regions of the moving groups.

This is likely a signature of mass segregation, which is a fairly rapid process and it can meaningfully change spatial distribution of stars in a cluster with respect to their mass in under 1 Myr \citep{allison2009}. Given that a binary system is going to be on average heavier than a single star, in equilibrating kinetic energies of all of the members, single stars are going to develop a larger amplitudes in their relative velocities, while multiple stars would end up sinking deeper into the potential well of their respective clusters. Such an evolution in the distribution of multiple stars has been previously observed in a number of open clusters \citep{sheikhi2016,jadhav2021,zwicker2024}, including some that have ages of just a few 10s of Myr. This is well matched with the timescales we observe for the rapid rotators to become more dominant in denser environments.

Furthermore, this is likely to be further amplified by the moving groups being kinematically coherent but in large part unbound with exception of the densest clusters housed in them. The population becomes tidally stretched, and the stars with most distinct relative velocities extending increasingly further away from the center of mass, with the tidal tails gradually decreasing in density. With larger velocity dispersion, single stars would be more likely to inhabit diffuse tidal tails, while binaries -- even if they are not bound to the cluster -- would typically trail closer to it.

As such, although initially binaries with separations of a few to a few tens of au appear to be less likely to form in denser parts of young populations than in the more diffuse parts, over time they should experience significant mass segregation in the timescales of a few tens of Myr which will redistribute them relative to the single stars.



\section{Conclusions}\label{sec:conclusions}

In this work, we have analyzed the fraction of rapid rotators compared to regular rotators in 6 different populations with ages of $<$10 Myr up to 150 Myr, binned into regimes of different density, from highly clustered regions, to the parts of the moving groups that are much more diffuse and unbound. These rapid rotators can be used as tracers of binary systems with separations of a few to a few tens of au.

We find that in younger populations, rapid rotators are more numerous in more diffuse environments. If they are binaries, denser clusters may be more likely to disrupt and dissolve these systems, producing a deficit of multiples in these clustered environments. This deficit may appear even at the final separations that are believed to be too tight to be affected by the cluster dynamics--at the earliest stages of formation they generally would form with wider separations, and over time they would settle into stable and tighter orbits.

On the other hand, as these populations evolve, at ages of $>$20 Myr, we observe that rapid rotators begin to be increasingly more likely to be found in denser regions. These populations will undergo mass segregation, which would end up redistributing single and binary stars throughout both the clusters and throughout the unbound moving group.

Given the biases involved in the selection of rapid rotators in populations of different ages, we are unable to compare the overall fraction of rapid rotators between disparate populations, and the overall conclusions are drawn only in a relative sense in each population individually. The observations of rapid rotators are suggestive of the above model of formation and evolution, however, we note that these results are preliminary. Follow up observations would be needed to explicitly confirm the candidate binaries through resolved observations, to better understand the separations they are sensitive to and to better understand the mass ratios of the systems involved. 



\software{TOPCAT \citep{topcat},\texttt{TESS\_localize} \citep{higgins2023}, \texttt{eleanor} \citep{feinstein2019}}

\begin{acknowledgements}
MK acknowledges support provided by NASA grant
80NSSC24K0620.

This work has made use of data from the European Space Agency (ESA)
mission {\it Gaia} (\url{https://www.cosmos.esa.int/gaia}), processed by
the {\it Gaia} Data Processing and Analysis Consortium (DPAC,
\url{https://www.cosmos.esa.int/web/gaia/dpac/consortium}). Funding
for the DPAC has been provided by national institutions, in particular
the institutions participating in the {\it Gaia} Multilateral Agreement.

\end{acknowledgements}

\bibliographystyle{aasjournal.bst}
\bibliography{main.bbl}

\end{document}